\documentclass[9pt,twocolumn,twoside]{osajnl}

\journal{ol} 

\setboolean{shortarticle}{true}

\usepackage{lineno}

\usepackage{amsmath}
\usepackage{epsfig}
\usepackage{graphicx}
\usepackage{epstopdf}
\usepackage{amssymb}
\usepackage{color,soul}
\usepackage{ulem}

\newcommand{\bs}{\boldsymbol}

\title{Tailoring on-axis spectral density with circularly coherent light beams}

\author[1,*]{O. Korotkova}
\author[2]{J. C. G. De Sande}
\author[3]{M. Santarsiero}
\author[4]{R. Mart\'inez-Herrero}
\author[4]{G. Piquero}
\author[3]{F. Gori}

\affil[1]{Department of Physics, University of Miami, 1320 Campo Sano Drive, Coral Gables, FL, 33146}
\affil[2]{ETSIS de Telecomunicación, Universidad Politécnica de Madrid, Campus Sur 28031 Madrid, Spain}
\affil[3]{Dipartimento di Ingegneria Industriale, Elettronica e Meccanica, Università Roma Tre, Via V. Volterra 62, 00146 Rome, Italy}
\affil[4]{Departamento de Óptica, Universidad Complutense de Madrid, Ciudad Universitaria, 28040 Madrid, Spain}

\affil[*]{Corresponding author: korotkova@physics.miami.edu
\newline
\newline
\bf © 2022 Optica Publishing Group. One print or electronic copy may be made for personal use only. Systematic reproduction and distribution, duplication of any material in this paper for a fee or for commercial purposes, or modifications of the content of this paper are prohibited.
\newline
\url{https://doi.org/10.1364/OL.458262}
}

\begin{abstract}
	The on-axis cross-spectral density (CSD) of a beam radiated by a stationary source with a circular coherence state and  a Gaussian spectral density is obtained in the closed form. It is revealed that the on-axis CSD is expressed via the Laplace transform of the source's degree of coherence or the Hilbert transform of the corresponding pseudo-mode weighting function. Such relations enable efficient tailoring of the on-axis spectral density, as we show with a slew of numerical examples.  
\end{abstract}

\setboolean{displaycopyright}{true}

\begin{document}
	
	\maketitle
	
	Structuring the spatial coherence states of stationary light sources makes it possible to efficiently control various statistical properties of radiated light beams, including spectral density, polarization properties, coherence state, scintillation index, etc.
	(\cite{Korotkova2021a}, Ch. 5). In particular, recently introduced and realized sources with circular coherence \cite{Santarsiero:OL17,Santarsiero:OL17b,new1, new2} have been shown to exhibit the self-focusing phenomenon, i.e., the formation of the on-axis intensity maximum in the propagating beam (c.f. \cite{Ding:JOSAA2017, app9132716, Sande:OE19, Cai:2020}). 
	The aim of this letter is to reveal a simple relation, with the help of the Laplace transform, existing between the on-axis cross-spectral density (CSD) and the source (circular) degree of coherence (DOC). Another relation between the on-axis CSD and the weighting function of the pseudo-modes, also conveniently used in describing stationary source' coherence states \cite{Gori:OL07},  relying on the Hilbert transform, is also derived. The application of either integral transform relation enables a straightforward calculation of the on-axis CSDs produced by a wide variety of circularly coherent sources. In this letter we restrict the analysis to tailoring of the on-axis spectral density, being the CSD at the coinciding arguments.   
	
	The CSD at transverse position vectors $\bs r_1, \bs r_2$ orthogonal to the $z$-axis and longitudinal positions $z_1$ and $z_2$ has form~\cite{ManWolf95}
	\begin{equation}
		\begin{split}	
			\displaystyle W(\bs r_1, z_1; \bs r_2, z_2) & =
			\displaystyle \int \int W_0(\bs \rho_1,\bs \rho_2) \\& \times  K_{z_1}^*(\bs r_1- \bs\rho_1) K_{z_2}(\bs r_2-\bs\rho_2) {\rm d}^2\rho_1 {\rm d}^2\rho_2 \; ,
		\end{split}
		\label{gen01}
	\end{equation}
	where $K_{z_j}$ denotes the free-space Green's function from $z=0$ to $z=z_j$  $(j=1,2)$ and $W_0(\bs \rho_1,\bs \rho_2)$ the CSD across the source. Within the validity of the paraxial approximation, the CSD along the $z$-axis ($ r_1 =  r_2 = 0$), denoted below by $W_z(z_1,z_2)$, takes form
	\begin{equation}
		\begin{split}
			\displaystyle W_z(z_1, z_2) & = \frac{e^{{\rm i} k(z_2 - z_1)}}{\lambda^2 z_1 z_2}  
			\displaystyle \int_0^{\infty}{\rm d}\rho_1 \int_0^{\infty} {\rm d}\rho_2 \int_0^{2\pi}{\rm d}\phi_1 \int_0^{2\pi}{\rm d}\phi_2\,  	\\& \times  
			\displaystyle  W_0(\bs \rho_1,\bs \rho_2) \rho_1 \rho_2 \exp\left[\frac{{\rm i}k}{2}\left(\frac{\rho_2^2}{z_2}-\frac{\rho_1^2}{z_1}  \right)  \right] 
			\; ,
		\end{split}
		\label{gen02}
	\end{equation}
	where $\lambda$ is the wavelength of the radiation, $k=2 \pi/\lambda$ and $(\rho,\phi)$ are the polar coordinates of vector $\bs{\rho}$. 
	
	For circularly coherent sources  \cite{Santarsiero:OL17} $W_0$ does not depend on the angular coordinates. In particular, for the Gaussian spectral density with the effective width $w_0$, $W_0$ becomes
	\begin{equation}
		\begin{array}{c}
			\displaystyle W_0(\rho_1,\rho_2)\propto \exp[-(\rho_1^2+\rho_2^2)/w_0^2] \,\mu(\rho_1, \rho_2)
			\; ,
		\end{array}
		\label{sou01}
	\end{equation}
	where $\mu$ is the DOC across the source \cite{ManWolf95}. Due to the circular symmetry the on-axis CSD in \eqref{gen02} takes on the form
	\begin{equation}
		\begin{split}
			\displaystyle W_z(z_1,z_2) & =\frac{k^2 e^{{\rm i}k(z_2 - z_1)}}{z_1z_2}  
			\int_0^{\infty}\int_0^{\infty} \exp\left[-\left(\frac{\rho_1^2+\rho_2^2}{w_0^2} \right)\right] 
			\\& \times
			\displaystyle \exp\left[\frac{{\rm i}k}{2}\left(\frac{\rho_2^2}{z_2} - \frac{\rho_1^2}{z_1}\right)\right]  \rho_1 \rho_2 \; \mu(\rho_1, \rho_2) \;  {\rm d}\rho_1 {\rm d}\rho_2
			\; ,
			\label{sou02}
		\end{split}
	\end{equation} 
	the proportionality factor of Eq. (\ref{sou01}) being omitted.
	
	As for the degree of coherence, we adopt a structure that specifically gives rise to circular coherence, viz.,

	\begin{equation}
		\begin{array}{c}
			\displaystyle \mu(\rho_1, \rho_2) =\mu\left(\frac{\rho_1^2-\rho_2^2}{\delta^2}\right)
			\; ,
			\label{sou03}
		\end{array}
	\end{equation} 
	with a real constant $\delta$. 
	%
	To simplify the integral in Eq.  (\ref{sou02}) we let
	\begin{equation}
		q = \left( \frac{w_0}{\delta}\right)^2\, ;
		\quad
		\displaystyle  \tau_j = \frac{\rho_j^2}{w_0^2} \, ; 
		\quad
		\displaystyle \zeta_j=\frac{z_j}{z_R} \, ;
		\quad 
		\sigma_{j}=1-\frac{{\rm i}}{\zeta_j}\, ; 
		\label{sou06b}
	\end{equation} 
	with $j=1,2$ and $z_R=\pi w_0^2/\lambda$. In such a way Eq.~(\ref{sou02}) yields
	\begin{equation}\label{CSD1}
		W_z(\zeta_1,\zeta_2)=K\int\limits_{0}^{\infty} \int\limits_{0}^{\infty}
		\exp[-\tau_1\sigma_1-\tau_2\sigma_2^*]
		\mu\left[q(\tau_1-\tau_2)\right]
		{\rm d}\tau_1 {\rm d} \tau_2,
	\end{equation}
	where
	\begin{equation} 
		K=(\zeta_1\zeta_2)^{-1}e^{2{\rm i}(\zeta_2-\zeta_1)z_R^2/w_0^2} .
	\end{equation}
	%
	This integral has form of the $q$-scaled 2D  Laplace transform with direct space variables $\tau_1$, $\tau_2$ and transform variables $\sigma_1$, $\sigma_2^*$:
	\begin{equation}\label{CSD11}
		W_z(\zeta_1,\zeta_2)=K\mathcal{L}[
		\mu(\sigma_1/q,\sigma_2^*/q)]
		/q^2.
	\end{equation}
	Then, using a relation between the 2D and 1D Laplace transforms (\cite{Ortigueira:Math20}, p.16), for the practically important case of real-valued $\mu$, we find that the on-axis CSD becomes
	\begin{equation}\label{WLT}
		W_z(\zeta_1,\zeta_2)=\frac{K}{q}\frac{\mathcal{L}[\mu(\sigma_1/q)]+\mathcal{L}[\mu(\sigma_2^*/q)]}{\sigma_1+\sigma_2^*}.
	\end{equation}
	We conclude that the CSD along the
	$z$-axis is directly related to the sum of the Laplace transforms of the degree of (circular) coherence calculated at $\sigma_1/q$ and $\sigma_2^*/q$.
	
	\begin{figure}
		\centering
		\includegraphics[width=0.8\linewidth]{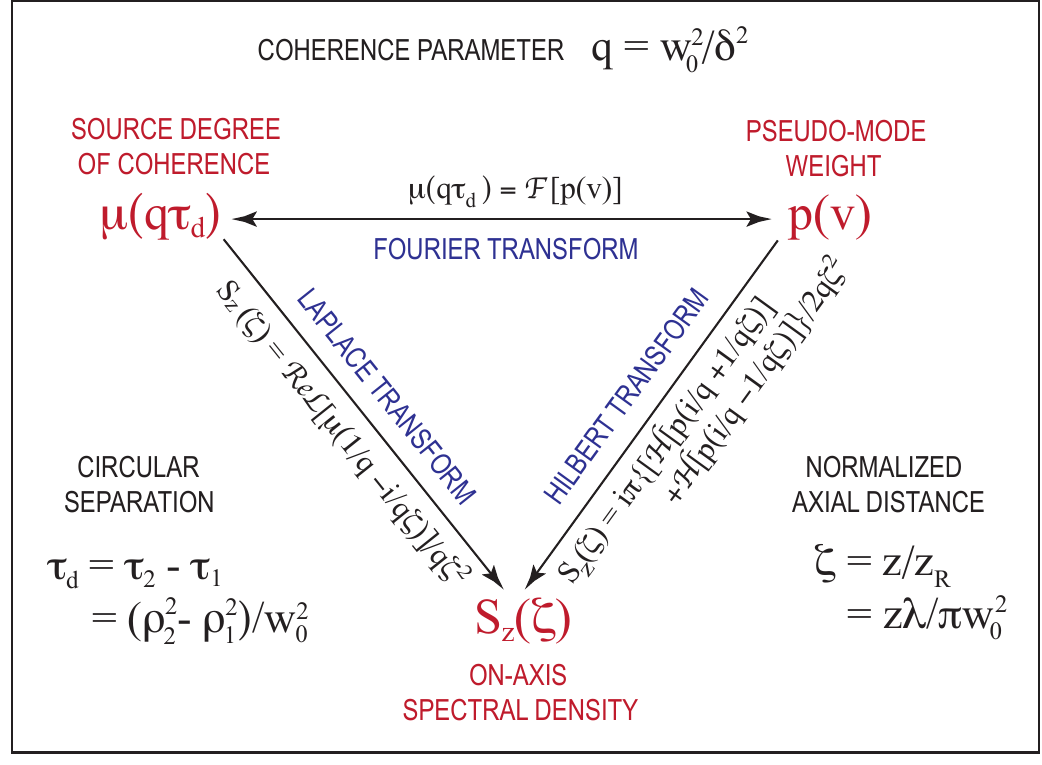}
		\caption{Relations for $\mu$, $p$ and $S_z$ for circularly coherent light.}
		\label{triangle}
	\end{figure}

	The on-axis spectral density $S_z(\zeta)=W_z(\zeta,\zeta)$ takes form
	\begin{equation}\label{VCZ}
		\begin{split}
			S_z(\zeta)&=
			\frac{1}{2q\zeta^2}\left\{\mathcal{L}\left[\mu \left(\frac{\zeta-{\rm i}}{q\zeta}\right)\right] + \mathcal{L}\left[\mu \left(\frac{\zeta+{\rm i}}{q\zeta}\right)\right]\right\}
			\\&=\frac{1}{q\zeta^2}\Re\mathcal{L}\left[ \mu\left( \frac{\zeta-{\rm i}}{q\zeta}\right) \right] \\&
			=\frac{1}{\zeta^2}\int\limits_0^{\infty} \mu(q\tau_{{ d}})\exp\left(-\tau_{ d}\right)\cos\left(\tau_{d}/\zeta \right){\rm d}\tau_{d},  
		\end{split}
	\end{equation}
	where $\tau_{ d}=\tau_1-\tau_2$. We  deduce at once that $S_z(\zeta)=S_z(-\zeta)$.  
	
	
	Structuring of the source DOC can also be made with the help of 
	the Parzen integral representation of CSD \cite{Gori:OL07,Parzen:AMS61}
	\begin{equation}
		\begin{array}{c}
			\displaystyle W_0(\bs \rho_1,\bs \rho_2) = \int  p(\bs v) H^*(\bs \rho_1,\bs v) H(\bs \rho_2,\bs v) {\rm d}^2v
			\; ,
		\end{array}
		\label{Bochner}
	\end{equation} 
	where $H$ is an arbitrary kernel and $p$ is a non-negative, Fourier transformable function. Indeed, this representation provides a simple sufficient condition ensuring that the CSD is \textit{ bona fide}.
	To obtain sources with circular coherence we set \cite{zcoh:OL2021}
	\begin{equation}\label{H}
		H(\rho)=\exp[-(b-{\rm i}v)\rho^2], \quad b>0.
	\end{equation}
	Substitution from \eqref{H} into \eqref{Bochner} yields the Fourier transform relation between $p(v)$ and $\mu(\tau_d)$, viz.,  $\mathcal{F}[p(v)]=\mu(\tau_d)$.
	
	\begin{table}[htbp]
		\centering
		\caption{\bf Examples of relations among $\mu(q\tau_d)$, $p(v)$, $S(\zeta)$.}
		\begin{tabular}{ccc}
			\hline
			$ \mu(q\tau_d)$ & $ p(v)$ & $  S(\zeta)$ \\ 
			\hline 
			 $ 1$ & $ \delta(v)$  & 
		     $  \frac{1}{1+\zeta^2}$
			\vspace{2px} \\ 
			$ \scriptstyle \exp(-q^2\tau_d^2) $ & $  \frac{\exp(-v^2/4q^2)}{2q\sqrt{\pi}} $ & $ \frac{\sqrt{\pi}}{2q\zeta^2}\Re\Biggl\{ \exp \left[ \frac{(\zeta-{\rm i})^2}{4q^2\zeta^2}\right] $ \vspace{2px} \\
			
				$\displaystyle $ & $\displaystyle $ & $ \times \text{erfc}\left( \frac{\zeta-{\rm i}}{2q\zeta}\right) \Biggr\} $ \vspace{3px} \\
				
			$ \cos{(q\tau_d)}$ & $ \frac{1}{2}\left[ \delta(v-q)+\delta(v+q)\right] $ & $ \frac{1 + (1 + q^2) \zeta^2}{1 - 2 (q^2-1) \zeta^2 + (1 + q^2)^2 \zeta^4} $ \vspace{5px} \\
			
			$ \frac{\sin(q\tau_d)}{q\tau_d} $ & $ \frac{1}{2q}\text{rect}(v/2q)$ & $ \frac{1}{q\zeta^2}\Re\left\{\arctan\left(\frac{q\zeta}{\zeta-{\rm i}}\right)\right\}$ \vspace{5px} \\
			
			
			$ J_0(q\tau_d)$ & $ \frac{\text{rect}(v/2q)}{\pi \sqrt{q^2-v^2}}$ & $ \frac{1}{\zeta^2}\Re\left\{\frac{1}{\sqrt{q^2+(1-{\rm i}/\zeta)^2}}\right\} $ \vspace{5px}\\
			$ \frac{2J_1(q\tau_d)}{q\tau_d}$ & $ \frac{2\text{step}\left(q^2-v^2\right)}{\pi q^2 (q^2-v^2)^{-1/2}}$ & $ \frac{2 \Re \left\{\sqrt{q^2+(1-{\rm i}/\zeta)^2}\right\}-2}{q^2\zeta^2} $ \vspace{2px} \\
			\hline
		\end{tabular}
		\label{tab:shape-functions}
	\end{table}
	
	We will now directly relate $p(v)$ and $W_z(\zeta_1,\zeta_2)$. Using the definition of the Laplace transform in formula (\ref{WLT}) and applying the Fourier-transform relation between $\mu$ and $p$ gives
	\begin{equation}
		\begin{split}
			W_z(\zeta_1,\zeta_2)&
		=\frac{K}{q(\sigma_1+\sigma_2^*)} \int\limits_0^{\infty}\int\limits_{-\infty}^{\infty} \left[e^{-\frac{\sigma_1}{q}\tau_d}+e^{-\frac{\sigma_2^{*}}{q}\tau_d} \right] e^{-{\rm i}v\tau_d}
		\\& 
		\times p(v){\rm d}v{\rm d}\tau_d.
		\end{split}
	\end{equation}
	 Next, changing the order of integrals and evaluating the one with respect to $\tau_d$ leads to expression
	\begin{equation}
		W_z(\zeta_1,\zeta_2)=\frac{K}{{\rm i}q(\sigma_1+\sigma_2^*)}\int\limits_{-\infty}^{\infty} \Biggl[  
		\frac{p(v)}{v-{\rm i}(\sigma_1/q)}  + 
		\frac{p(v)}{v-{\rm i}(\sigma_2^*/q)}\Biggr] {\rm d}v.
	\end{equation}
	On recognizing the Hilbert transform defined as \cite{Bracewell00} 
	\begin{equation}
		\mathcal{H}[f(t)]= \frac{1}{\pi} 
		\int\limits_{-\infty}^{\infty} \frac{f(t')}{t-t'}{\rm d}t',     
	\end{equation}
	where the divergence at $t=t'$ is allowed for by taking the Cauchy principal value of the integral, 
	we express $W_z$ as  
	\begin{equation}\label{HilbertW}
		W_z(\zeta_1,\zeta_2)=\frac{{\rm i} K\pi}{q(\sigma_1+\sigma_2^*)}\left\{  
		\mathcal{H}[p({\rm i}\sigma_1/q)]+ \mathcal{H} [p({\rm i} \sigma_2^*/q)]
		\right\}.
	\end{equation}
	Hence the spectral density reduces to expression
	\begin{equation}\label{HilbertS}
		S_z(\zeta)=\frac{{\rm i}\pi}{2q\zeta^2}\left\{
		\mathcal{H}\left[p\left(\frac{{\rm i} \zeta+1}{q\zeta}\right)\right]+\mathcal{H} \left[p\left(\frac{{\rm i}\zeta-1}{q\zeta}\right)\right]
		\right\}\, .
	\end{equation}
	Similarly to \eqref{VCZ}, the direct relation between $p(v)$ and $S_z(\zeta)$ in \eqref{HilbertS} provides with a convenient tool for  the on-axis spectral density tailoring. Equations (\ref{VCZ}) and (\ref{HilbertS}) are the main analytic results of the letter. They are summarized in Fig. \ref{triangle}.  
	
	\begin{figure}[htbp]
	\centering
	\includegraphics[width=0.98\linewidth]{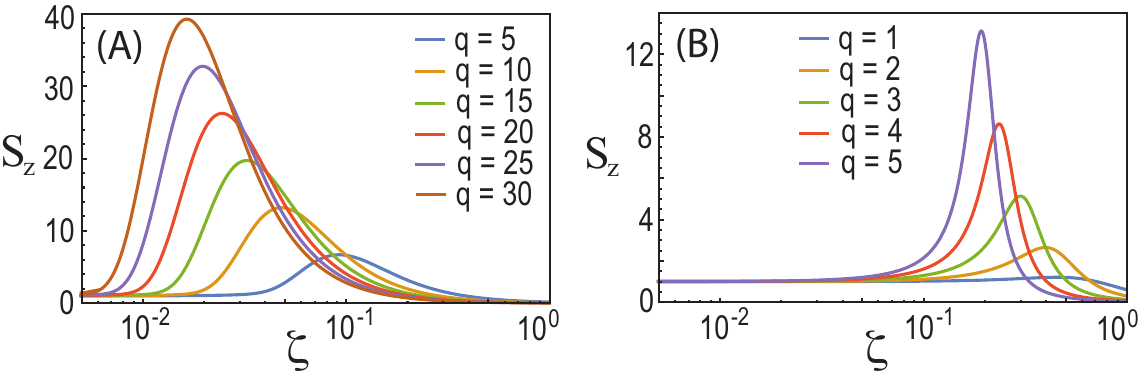}
	\caption{Spectral density $S(\zeta)$ of a beam (A) with Gaussian DOC; 
		(B) with cosine DOC, for several $q$ values.} 
	\label{fig1}
\end{figure}

	A list of important one-parameter beam families with circular coherence is provided in Table \ref{tab:shape-functions} with specification of $\mu(q\tau_d)$, $p(v)$ and $S_z(\zeta)$. The first line of the table corresponds to the limiting case of a completely coherent source. In this case,  either calculating the Laplace transform of unity, $\mathcal{L}[1]=1/t$, in \eqref{VCZ}, or using the property of Hilbert transform  $\mathcal{H}[\delta(t)]=1/(\pi t)$, results, after algebra, in the Lorentzian profile, i.e.,
	\begin{equation}
		S_z(\zeta)=(1+\zeta^2)^{-1}.
	\end{equation}
	
	Line 2 of the table involves the Gaussian $\mu$-$p$ pair,
	\begin{equation}
	\mu(q\tau_d)=\exp(-q^2\tau_d^2), \quad  p(v)=\frac{1}{2q\sqrt{\pi}}\exp\left(-\frac{v^2}{4q^2}\right). 
	\end{equation}
	Either of the transforms returns, for $S_z$, the expression	\begin{equation}\label{WLT1}
		S_z(\zeta)=\frac{\sqrt{\pi}}{2q\zeta^2}\Re\Biggl\{ \exp \left[\frac{(\zeta-{\rm i})^2}{4q^2\zeta^2}\right] \text{erfc}\left( \frac{\zeta-{\rm i}}{2q\zeta}\right) \Biggr\},
	\end{equation}
	being consistent with a previous result~\cite{zcoh:OL2021}. While the application of the Laplace transform $\mathcal{L}[\mu(q^2\tau_d^2))]=\sqrt{\pi}\exp(t^2/4q^2)\text{erfc}(t/2q)$ is straightforward, the Hilbert transform formula in \eqref{HilbertS} can be viewed as combination of Faddeeva functions defined as \cite{Faddeeva}
	\begin{equation}
		\begin{split}
			w(s)=\frac{{\rm i}}{\pi} \int\limits_{-\infty}^{\infty} \frac{e^{-t^2}}{s-t}{\rm d}t, 
		\end{split}
	\end{equation}
	having representation $w(s)=\exp(-s^2)\text{erfc}(-{\rm i}s)$ for complex arguments. Thus, with $s=\sigma/2q$ and $\sigma^*/2q$ in the first and second transforms, \eqref{HilbertS} returns the same result as in \eqref{WLT1}. 
	
Line 3 in Table \ref{tab:shape-functions} includes pair
	\begin{equation}
	\mu(q\tau_d)=\cos(q\tau_d), \quad p(v)=  \left[\delta(v+q) + \delta(v-q) \right]/2 \, . 
	\end{equation} 
	Using again the Hilbert transform of the $\delta (t)$ function and its  shifting property, the spectral density can be found as 
		\begin{equation}\label{Sz_cos}
		S_z(\zeta)=\frac{1 + (1 + q^2) \zeta^2}{1 - 2 (q^2-1) \zeta^2 + (1 + q^2)^2 \zeta^4}
		\, .
		\end{equation}	

Figure \ref{fig1} (A) and (B) shows $S_z(\zeta)$ for the Gaussian and the cosine DOC, given in lines 3 and 4 of Table \ref{tab:shape-functions}, respectively. In this and the all the figures below the log-linear scale is used for making it possible to better  distinguish the $S_z$ profiles at small values of $\zeta$. Since these families are single-parametric the amount of self-focusing also determines the location of focus. 

\begin{figure}
	\includegraphics[width=0.98\linewidth]{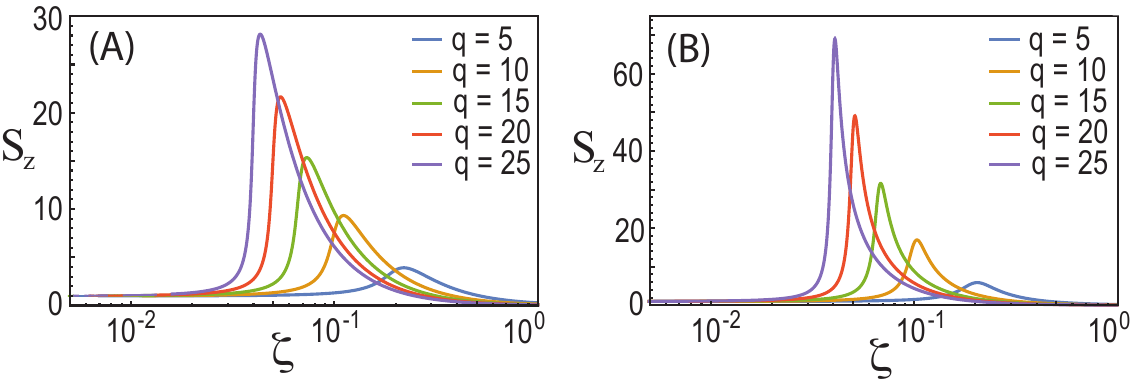}
	\caption{Spectral density $S_z(\zeta)$ of a beam with (A) sinc DOC; (B) $J_0$-Bessel DOC, for several values of $q$.
	}
	\label{fig2}
\end{figure}
	\begin{figure}
		\centering
		\includegraphics[width=0.98\linewidth]{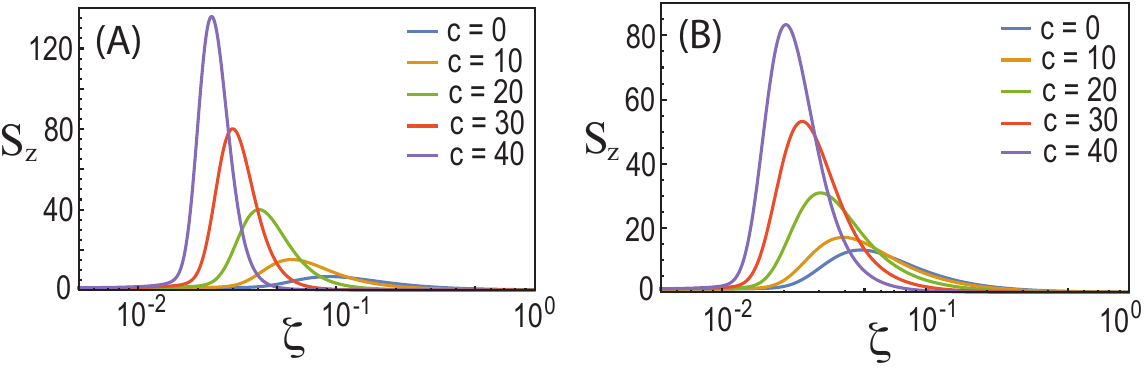}
		\caption{Spectral density $S(\zeta)$ of a beam with cos-Gaussian DOC. (A) $q=5$; (B): $q=10$.} 
		\label{fig3}
	\end{figure}


	%
	%
	
	Lines 4-6 in Table \ref{tab:shape-functions} present other examples of sources, for which $S_z$ may be conveniently derived from formulas (\ref{WLT}) or (\ref{HilbertS}). In particular, in line 5 the $\mu-p$ pair 
	\begin{equation}
		\mu(q\tau_d)=\frac{\sin(q\tau_d)}{q\tau_d}, \quad p(v)= \frac{1}{2q} {\rm rect}\left(\frac{v}{2q}\right) \, . 
	\end{equation} 
	where ${\rm rect}(v)=1$ if $|v|<1/2$ and zero otherwise. For this case the on-axis spectral density results in expression
	\begin{equation}\label{Sinc}
		S_z(\zeta)=\frac{\sqrt{\pi}}{2q\zeta^2}\Re \left[\arctan\left( \frac{q \zeta}{\zeta-{\rm i}}\right)  \right]
		\, ,
	\end{equation}

The last two lines of the table present the analytic results for $S_z$ for the source with a DOC involving the $J-$Bessel functions. The latter model sources are known as the besinc circularly correlated sources \cite{Santarsiero:OL17b}. Figure \ref{fig2} shows $S_z(\zeta)$ plotted for two of these DOC's with several values of $q$. A sharp increase in spectral density is observed before its maximum is reached.

	
	
	More flexibility in tailoring the on-axis spectral density could be gained if we consider a two-parameter DOC. This can be generally achieved by employing products/convolutions of $\mu(q\tau_d)$ or $p(v)$. For example, one may modulate the DOC in the Gaussian case by a cosine function, i.e.,
	\begin{equation}\label{cosG}
		\mu(q\tau_d,c)=\exp(-q^2\tau_d^2)\cos[(c/q)\tau_d].
	\end{equation}
	The corresponding $p(v)$ becomes
	\begin{equation}
	p(v)= \exp\left( - v^2/4q^2\right) \cosh[(q/c)v],   
	\end{equation}
being non-negative. The cosine term can be expressed by the Euler's formula, $\cos(s)=0.5[\exp({\rm i} s)+\exp(-{\rm i} s)]$, 
	and the exponential scaling property of the Laplace transform can then be applied, resulting in the spectral density 
	%
	\begin{equation}\label{SWLT}
		\begin{split}
			S_z(\zeta)&
			=
			\frac{\sqrt{\pi}}{4q \zeta^2}\Re \Biggl\{ \exp\left[\left( \frac{\zeta- {\rm i}-{\rm i} c}{2q\zeta}\right)^2 \right] \text{erfc}\left(\frac{\zeta- {\rm i}-{\rm i} c}{2q\zeta}\right)  \Biggr\} 
			\\ & + 
			\frac{\sqrt{\pi}}{4q \zeta^2}\Re \Biggl\{ \exp\left[\left( \frac{\zeta- {\rm i}+{\rm i} c}{2q\zeta}\right)^2 \right] \text{erfc}\left(\frac{\zeta- {\rm i}+{\rm i} c}{2q\zeta}\right)  \Biggr\}  \, .
		\end{split}
	\end{equation}
	
	Figure \ref{fig3} shows spectral density $S_z(\zeta)$ of the beam with circular cos-Gaussian DOC. As parameter $c$ grows exceptional focusing ability is illustrated, while the location of the maximum is still kept away from the source plane. 
	
	So far, all the considered examples showed either a monotonic decay (for coherent case) or a single maximum of $S_z(\zeta)$ whose position and height were tuned by the parameters of the source DOC. A finer control of $S_z(\zeta)$ can be achieved by using legitimate linear combinations of the previously considered DOCs, while retaining the Gaussian profile of width $w_0$ for the source spectral density, as given in \eqref{sou01}. Indeed, consider  
	%
	%
	
	%
	\begin{equation}
		\begin{array}{c}
			\displaystyle \mu(q\tau_d,M) =\frac{1}{\mu_0(M)} \sum_{m=1}^{M} A_m \exp{\left[ - q_m^2\tau_d^2  \right] } \; .
			\label{mu_sup}
		\end{array}
	\end{equation} 
	Here $A_m>0$ is the weighting coefficient and $q_m$ is the correlation width of term $m$, respectively, while  $\mu_0(M)=\mu(0,M)$ is the normalization factor. Then, substitution from  \eqref{mu_sup} into \eqref{WLT} yields the axial spectral density $S_z$ in form
	%
	%
	\begin{equation} \label{S_sup}
		\displaystyle S_z(\zeta) = \frac{\sqrt{\pi}}{2\mu_0(M)\zeta^2} \sum_{m=1}^{M} \frac{A_m}{q_m}
		\Re \left\{ \exp\left[\frac{(\zeta-{\rm i})^2}{4q_m^2\zeta^2}\right] \text{erfc} \left( \frac{\zeta-{\rm i}}{2q_m\zeta }\right)\right\} \, .
	\end{equation} 

\begin{figure}
		\centering
		\includegraphics[width=0.98\linewidth]{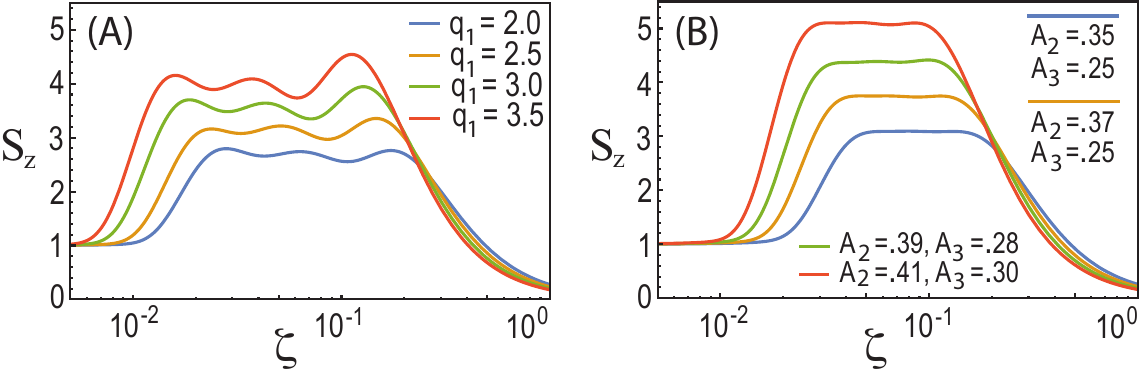}
		\caption{Spectral density $S_z(\zeta)$, plotted from \eqref{S_sup} with $M=3$, $A_1=1$.
		The rest of parameters are: (A) $q_3=3q_2=9q_1$, $A_2=0.2$, $A_3=0.1$; 
		(B) $q_3=2.5q_2=5q_1$, $q_1$ same as in (A).
		}
		\label{fig:gc3}
	\end{figure}

	Figure \ref{fig:gc3} shows the 
	possibilities stemming from the use of model \eqref{mu_sup} having three terms with specific values of $A_m$ and $q_m$. Figure  \ref{fig:gc3}(A) shows the $S_z$ curves with three maxima and Fig. \ref{fig:gc3}(B) demonstrates a region of a nearly constant $S_z$. 
		\begin{figure}
		\centering
		\includegraphics[width=0.98\linewidth]{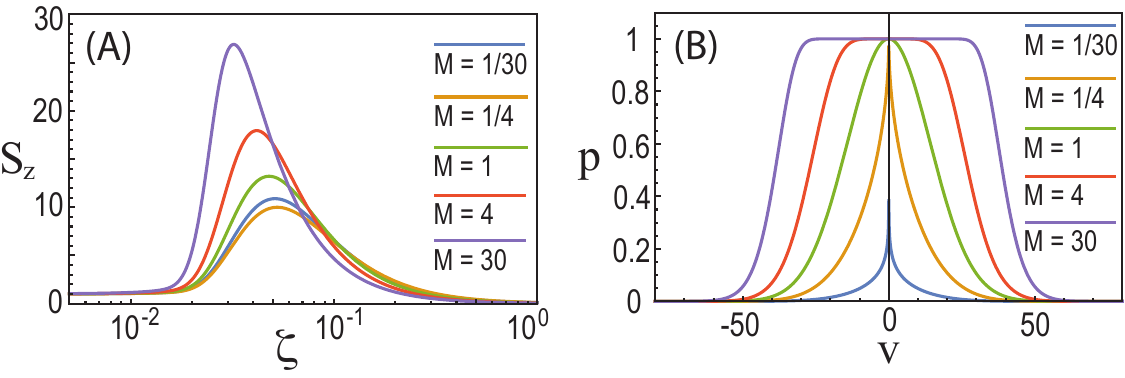}
		\caption{Circular MG beams with $q=10$: (A) $S_z(\zeta)$; (B) $p(v)$. 
		}
		\label{fig:MG}
	\end{figure}
	
	
	In this connection it is instructive to analyze extension of circular Gaussian DOC to the Multi-Gaussian (MG) family~\cite{Korotkova:JOSAA12}:
	\begin{equation}\label{muMG}
		\mu(q\tau_d,M)=\frac{1}{\mu_0(M)}\sum\limits_{m=1}^{\infty}\frac{(-1)^{m+1}(M)_m}{mm!}\exp\left[-q^2_m \tau_d^2\right], 
	\end{equation}
	where $M>0$, $\mu_0(M)=\mu(0,M)$ is the  normalization factor, $q_m=q/\sqrt{m}$ and  $(M)_m=M(M-1)...(M-m+1)$
	is the Pochhammer symbol. The corresponding $p(v)$ takes form  
	\begin{equation}
			p(v)=\sum\limits_{m=1}^{\infty}\frac{(-1)^{m+1}(M)_m}{m!}\exp(-m v^2/4q^2),
	\end{equation}
	and non-negative definiteness is ensured: $p(v)=1-(1-\exp[-v^2/4q^2])^M>0$. Using \eqref{muMG} in \eqref{WLT} yields 
	\begin{equation}\label{MWLT}
		\begin{split}
			S_z(\zeta) & = \frac{\sqrt{\pi}}{2\mu_0(M)\zeta^2} \sum\limits_{m=1}^{\infty}\frac{(-1)^{m+1}(M)_m}{mm!q_m} \\& \times
			\Re \left\{ \exp\left[\frac{(\zeta-{\rm i})^2}{4q_m^2\zeta^2}\right] \text{erfc} \left( \frac{\zeta-{\rm i}}{2q_m\zeta }\right)\right\}.
		\end{split}
	\end{equation}
	
	Figure \ref{fig:MG}(A) shows $S(\zeta)$ corresponding to the circular MG family, for several values of $M$. While case $M=1$ corresponds to circular Gaussian DOC discussed above, those for $M>1$ ($M<1)$ enable enhancement (suppression) of self-focusing. The maximum also shifts towards (away) from the origin for $M>1$ ($M<1$). Importantly, with the increase of $M$, the self-focusing effect saturates at about a factor of 2 for $M>1$ and 0.7 for $M<1$. Thus, the MG circularly coherent sources also make it possible to control the location/amount of focusing separately, just like circular cosine-Gaussian correlated sources in \eqref{cosG}. However, regardless of the $p$-profile shown in Fig. \ref{fig:MG}(B), the evolution of $S_z$ does not involve a flat/cusp area. This is due to the fact that $S_z$ is not in the Fourier transform relation with $\mu$ and, hence, reconstruction of the $p$-profile must not be expected.

	To summarize, we have expressed the spectral density $S_z$ of a circularly coherent beam via Laplace transform of the source degree of coherence $\mu$ and via Hilbert transform of the pseudo-mode weighting function $p$. Together with the Fourier transform relation between $\mu$ and $p$, our  new relations provide a convenient analytic tool for design of stationary sources with prescribed axial dynamics of the spectral density. Our findings will benefit any application involving directed energy carried by random light fields, such as profilometry \cite{Rosen:AO00}, optical coherence tomography \cite{Ahmad:OL19} and quantitative phase microscopy \cite{PopescuBook}.   
	
	\begin{backmatter}
		\bmsection{Funding} Spanish Ministerio de Econom\'ia y Competitividad, project PID2019-104268 GB-C21. 
	
		\bmsection{Disclosures} The authors declare no conflicts of interest.
		
		\bmsection{Data availability} No data were generated or analyzed in the presented research.
		
		
	\end{backmatter}

%
%
%

\end{document}